\documentclass[pre,twocolumn, showpacs]{revtex4}
\usepackage{graphicx}
\usepackage{dcolumn}

\begin{document}

\title{Semiflexible Polymers in a Random Environment}
\author{Arti Dua}
\affiliation{Max-Planck-Institute for Polymer Research - P.O. Box 3148,
  D-55021 Mainz, Germany}
\author{Thomas A. Vilgis}
\affiliation{Max-Planck-Institute for Polymer Research - P.O. Box 3148,
  D-55021 Mainz, Germany}
\date{\today}

\begin{abstract}
  We present using simple scaling arguments and one step replica symmetry
  breaking a theory for the localization of semiflexible polymers in a
  quenched random environment. In contrast to completely flexible polymers,
  localization of semiflexible polymers depends not only on the details of the
  disorder but also on the ease with which polymers can bend. The interplay of
  these two effects can lead to the delocalization of a localized polymer with
  an increase in either the disorder density or the stiffness. Our theory
  provides a general criterion for the delocalization of polymers with varying
  degrees of flexibility and allows us to propose a phase diagram for the
  highly folded (localized) states of semiflexible polymers as a function of
  the disorder strength and chain rigidity.
\end{abstract}

\pacs{36.20.-r Macromolecules and polymer molecules - 05.40.-a
  Fluctuation phenomena, random processes, noise and Brownian motion - 
  75.10.Nr Spin-glass and other random models - 71.55.Jv Disordered
  structures; amorphous and glassy solids} 

\maketitle

\section{Introduction}

In problems concerning flexible polymers in disordered media, disorder seems
to play a more important role than the conformational properties of the
polymers [1-12]. The effect is usually manifested in the size of the polymer,
which in a simple but rather unrealistic case of Gaussian chains in a random
potential, is found to be independent of the chain length $L$. The radius of
gyration in three dimensions simply scales as $R \simeq
b/(\tilde{\Delta}/b^{3})$, where $b$ is a typical cut off length and plays the
(undetermined) role of the segment size, and $\tilde{\Delta}$ is the disorder
strength at a given system volume.  The problem, however, is academic mainly
on two accounts: firstly, the effects due to excluded volume interactions are
completely ignored, and secondly the effects of finite extensibility are not
captured by the infinitely flexible Gaussian chains. It is conceivable that
the inclusion of these effects, though nontrivial, can modify our present
understanding of the problem.

An effort in this direction has recently been made by accounting for the
effects of excluded volume interactions \cite{Gold2}. The results show that a
chain undergoes a delocalization transition as the strength of the
self-avoiding interaction is increased. Since localization is usually
considered as filling up of the entropic traps --- created by randomness ---
with polymeric materials, it is understandable that a self-avoiding chain can
fill these holes up to a density close to $\rho b^{3} \sim {\cal
  O}(1)$ ($\rho$ is the monomer density) before
delocalizing and moving to the next hole. For Gaussian chains, on the
contrary, delocalization does not occur simply because they have an infinite
stored length and no limit up to which the holes can be filled.

In this paper we address the second of the above mentioned problems, i.e.,
semiflexible polymers in a random environment. Since semiflexible polymers are
finitely extensible, the theory rectifies the infinitely flexible Gaussian
chain model and serves as a useful description for localization of chains with
varying degrees of flexibility. In contrast to Gaussian chains, localization
of semiflexible polymers depends rather strongly on the conformational
properties of the polymer, i.e., the ease with which a chain can bend and
eventually adopt highly packed folded states. While a polymer localizes at a
certain fixed strength of the disorder, it can get delocalized with an
increase in the stiffness which can be brought about by decrease in the
temperature. A chain of a given fixed stiffness, on the other hand, can get
delocalized with the increase in the disorder strength. The interplay of the
disorder strength and bending energy is instrumental in bringing about
localization of semiflexible polymers.
 
The following section reviews the salient features of the wormlike
chain (WLC) model; it provides a general criterion for localization of a
semiflexible chain in random environment using simple scaling
arguments. Section III summarizes the analytical difficulties involved
in treating the WLC model and discusses the Weiner model for stiff
chains. Section IV presents a rigorous analysis of semiflexible
chains in random media using the one step replica symmetry breaking
calculations. The results are summarized in Section V. The details of the
one step replica symmetry calculations are included in Appendix A.

\section{Model and Scaling}

The configurational statistics of a semiflexible polymer is described by the
wormlike chain model \cite{yama, saito}. In this model, a chain is considered as a continuous
curve of total length $L$ parameterized by an arc length variable $s$. The
variation of the unit tangent vector ${\bf u}(s) = \partial{\bf
  r}(s)/\partial{s}$ at different points along the curve with respect to the
arc length variable determines the energy cost of bending [15-24]. To the
lowest scalar order in $\partial{\bf u}(s)/\partial{s}$, the bending energy is
given by
\begin{equation}
\beta{\cal H} = 
{l_{p}\over 2}\int_{0}^{L}ds\left( \frac{\partial {\bf u}}{\partial s}
\right)^{2},
\end{equation}
where $\beta$ is $(k_{\rm B}T)^{-1}$ and $l_{p}$ is the
persistence length of the polymer. It is a measure of the distance
over which bond vectors are correlated, and hence determines the
effective stiffness of the chain. In general, the correlation between
the unit tangent vectors at two different points along the chain
backbone decays exponentially, and is given by 
\begin{equation}
\langle {\bf u}(s)\cdot {\bf u}(s')\rangle = \exp(-|s-s'|/l_{p}).
\end{equation}
The persistence length is related to the bending energy $\epsilon$ by
$l_{p} = \beta \epsilon$. If such a chain interacts with $N$ randomly
distributed obstacles, the Hamiltonian is of the form
\begin{equation}
\label{hamilt}
\label{origH}
{\beta \cal H} = 
{l_{p}\over 2}\int_{0}^{L}ds \left( \frac{\partial {\bf u}}{\partial s}
\right)^{2} + \sum_{i=1}^{N}\int_{0}^{L}ds W[{\bf r}(s)-{\bf r}_{i}],
\end{equation}
where ${\bf r}(s)$ is the position vector of a segment at distance $s$ from
one end of the chain; and ${\bf r}_{i}$ is the position vector of the $i$th
obstacle. $W$ is the potential that describes the interaction between the
polymer and the obstacles, the form of which is yet to be specified. If the
obstacle density ${\rho({\bf r})}$ is defined by
\begin{equation} 
\rho({\bf r}) = \sum_{i=1}^{N} \delta({\bf r} - {\bf r}_{i}),
\end{equation}
the potential can simply be written as
\begin{equation}
\label{potential}
 \sum_{i=1}^{N} W[{\bf r}(s)-{\bf r}_{i}] = \int d{\bf
   r} W[{\bf r}(s)-{\bf r}] \rho({\bf r}). 
\end{equation}
At the simplest level of description, the obstacle density can be
assumed to have a Gaussian distribution \cite{Edwards,muth}:
\begin{equation}
P[\rho({\bf r})] = \exp\left(- \frac{1}{2} \int d{\bf r} 
\frac{ (\rho({\bf r}) - \rho_{0})^2}{\rho_{0}}\right),
\end{equation}
where $\rho_{0}$ is the mean obstacle density given by
\begin{equation}
\rho_{0} = \frac{1}{V}  \int d{\bf r}\rho({\bf r}). 
\end{equation}
To describe the effect of the disorder, one needs to perform an average over the obstacle density. Since the average is quenched, the estimation of the free energy relies on a difficult task of computing the average of the logarithm of the partition function $Z$. The problem is usually circumvented by introducing $n$ independent replicas of the initial system such that
\begin{equation}
\ln Z = \lim_{n \rightarrow 0} (Z^n -1) /n.
\end{equation} 
The replicated partition function can easily be averaged over the
Gaussian random potential given by Eq. (6) to yield
\begin{equation}
\left< Z^{n} \right> = \int_{{\bf r}(0)=0}^{{\bf r}(L)=R}{\cal{D}}[{\bf
  r}(s)] \exp[- \beta {\cal H}_{n}], 
\end{equation}
where the effective Hamiltonian is given by
\begin{eqnarray}
{\beta {\cal H}_{n}} &=& {l_{p}\over 2}\sum_{\alpha = 1}^{n}\int_{0}^{L}ds \left( \frac{\partial
    {\bf u_{\alpha}(s)}}{\partial s} \right)^{2} \nonumber\\ &-&
\frac{\rho_{0}}{2}\sum_{\alpha, \beta = 1}^{n}\int_{0}^{L}ds \int_{0}^{L} ds^{\prime} U[{\bf r_{\alpha}}(s)-{\bf r_{\beta}}(s^{\prime})], 
\end{eqnarray}
where 
\begin{equation}
U[{\bf r}_{\alpha}(s)-{\bf r}_{\beta}(s^{\prime})] = \int d{\bf r} W[{\bf r_{\alpha}}(s)-{\bf
  r}] W[{\bf r_{\beta}}(s^{\prime}) - {\bf r}]
\end{equation}
If the interaction between the polymer and the obstacles is assumed to be
short ranged, i.e., $W[{\bf r}] \propto \delta({\bf r})$, then the potential
can be expressed in terms of the natural scales, i.e., $U({\bf r})\equiv gl_{p}^4\delta({\bf r})$.  In writing the last expression for $U({\bf r})$, we have introduced a dimensionless coupling constant $g$, which in general contains microscopic details. The natural (short) lengthscale defined in the problem is the persistence length $l_{p}$ itself. Therefore, we can  express the scaling of the potential in terms of the persistence length; this makes the second term of Eq. (10) dimensionless, as should be the case \cite{Edwards}.  Having substituted this form of the potential, the effective Hamiltonian is given by
\begin{eqnarray}
\label{annealed}
{\beta {\cal H}_{n}} &=& {l_{p}\over 2}\sum_{\alpha = 1}^{n}\int_{0}^{L}ds \left( \frac{\partial
    {\bf u_{\alpha}(s)}}{\partial s} \right)^{2} \nonumber\\ &-&
\frac{\Delta}{2} \sum_{\alpha, \beta = 1}^{n}\int_{0}^{L}ds \int_{0}^{L} ds^{\prime} \delta[{\bf r_{\alpha}}(s)-{\bf r_{\beta}}(s^{\prime})],
\end{eqnarray}
where $\Delta = gl_{p}^4\rho_{0}$ and has the dimensions of length. Formally the second term in the above expression can be interpreted as a random potential which acts along the contour of the chain; it is a starting point of many studies on polymers in a disordered environment (see e.g. \cite{Miglior}). In the present study, the use of an array of obstacles instead of a random potential introduces a  natural length scale in terms of the mean distance between the obstacles, $\l_{\rm obs} \simeq 1/\rho_{0}^{1/3}$. It turns out to be useful since this length scale can directly be compared with the persistence length in various physical situations. 

Let us proceed by providing a rough estimate of the different energies
involved using simple scaling arguments. This can be done by dimensional analysis of the above Hamiltonian. In the case of annealed average there is no coupling between various replicas, that is $\alpha = \beta$, and the replica trick is not necessary. The free energy in three dimensions scales as
\begin{equation}
\beta \left<{\cal F}\right> \simeq \frac{l_{p}L}{R^{2}} - \Delta {L^{2}\over R^{3}}.
\end{equation}
The first term accounts for the bending energy penalty (per $k_{\rm B}T$) for
a chain of length $L$. It is typically defined as the ratio of the bending
force constant $\beta \epsilon$ (= $l_{p}$) to the radius of curvature $R$.
Note that the first term in Eq. (\ref{annealed}) corresponds to the
bending energy per thermal energy of a semiflexible chain. The derivative of
the tangent vector $\bf{u}(s)$ is nothing but the local curvature. The second
term corresponds to excluded volume type interactions between the chain segments, which by virtue of the disorder average, are effectively attractive. In the annealed case the problem is similar to a semiflexible chain with attractive interactions. 


In the case of quenched disorder, on the other hand, there is a coupling between various replicas and the free energy scales as 
\begin{equation}
\label{quenched0}
\beta \left<{\cal F}\right> \simeq \frac{l_{p}L}{R^{2}} - \tilde{\Delta}^{1/2} \frac{L}{R^{3/2}},
\end{equation}
where the second term in Eq. (\ref{quenched0}) describes a typical energy
barrier of the quenched random potential; it is usually estimated as the root
of the variance using simple dimensional analysis \cite{Natter,Miglior}. It is
to be noted that $\tilde{\Delta} = \Delta{|\ln V|}$, where $V$ is the volume
of system. The issue of finite system volume is important for the case of
quenched disorder since in an infinite random medium quenched and annealed
disorders lead to the same results \cite{Cates,Gold1}. In what follows it is
assumed that the system volume is much larger than the polymer dimensions
$R^3$.  The minimum of quenched free energy estimate yields the following
result for the size of the chain,
\begin{equation}
\label{locsize1}
R_{\rm loc} \simeq \l_{p} \left( \frac{l_{p}}{\tilde{\Delta}} \right),
\end{equation}
which, as in the limit of totally flexible chains, does not depend on the chain
length itself but is completely determined by the disorder and the
persistence length.  Although the size turns out to be similar to
that of a Gaussian chain, its dependence on $l_{p}$ instead of $b$ suggests a
natural condition on the dimensionless ratio $l_{p}/\tilde{\Delta}$. That this
indeed is the case can be seen from the following arguments: when the chain is
reasonably stiff $l_{p}\approx L$ and $R_{\rm loc}$ can at most be of the
order of $L$, which immediately implies that $l_{p}/L \leq
\tilde{\Delta}/l_{p} $; in the limit of sufficiently flexible chains, it is
expected that $R_{\rm loc}\leq R_{\rm gauss}(\approx \sqrt{Ll_{p}})$, which
implies $(l_{p}/L)^{1/2}\leq \tilde{\Delta}/l_{p}$.

In addition to the upper cut-off on the size $R_{\rm loc}$, there is a lower
cut-off, which is important for a flexible chain and suggests that $R_{\rm loc}$ can not be less than the
persistence length $l_{p}$. $R_{\rm loc} \geq l_{p}$ therefore implies
$\tilde{\Delta}/l_{p} \leq 1$. For a flexible chain, i.e., $L/l_{p} > 1$, the
latter constraint together with the constraint $l_{p}/L \leq
\tilde{\Delta}/l_{p} $ can be summarized as
\begin{equation}
\label{cond1}     
\left( \frac{l_{p}}{L} \right)^{1/2} \leq \frac{\tilde{\Delta}}{l_{p}} \leq 1.
\end{equation}
The second inequality implies that $l_{p}< (g \rho_{0})^{-1/3}$, which
suggests that as the chain become stiffer it localizes at progressively lower
density of the disorder $\rho_{0}$. In other words, since $l_{\rm obs} \approx
\rho_{0}^{-1/3}$ the chain is localized as long as its persistence length
$l_{p}$ is less than the mean separation between the obstacles $l_{\rm obs}$.
It is to be noted that the size of the chain has a simple
interpretation in terms of the obstacle density. This can be seen by substituting 
$\Delta \propto l_{p}^{4}\rho_{0}$ into Eq. (\ref{locsize1}) to produce $R_{\rm loc} \propto l_{p} (l_{\rm
  obs}/l_{p})^{3}$. Then the criterion for localization, as discussed above, is simply given by the
comparison between the persistence length and the mean distance between the obstacles. 

As discussed in the Introduction, a Gaussian chain undergoes a
continuous decrease in its size with an increase in the obstacle
density since the chain can be packed without limit into the cavities
formed by the obstacles. In the presence of excluded volume
interactions, however,  a Gaussian chain can fill this cavity only up
to a point before delocalizing and moving on to the next cavity; at
high obstacle density the chain eventually localizes by forming pearl
necklaces (as discussed in Ref. \cite{Gold2}). The present theory
shows that a semiflexible chain \emph{without} excluded volume
interactions can delocalize when the obstacle density and persistence
length are related by Eq. (\ref{cond1}). This is in marked constrast
to the infinitely flexible Gaussian chain which can delocalize only
when aided by the excluded volume interaction. In what follows we
shall study the localization-delocalization transition of a
semiflexible chain in the absence of excluded volume interactions,
paying special attention to the flexible limit ($L/l_{p} > 1$). We
will show that even in the flexible limit, new physics arises because
the persistence length defines a lower cut-off length scale which is
absent in the Gaussian chain. We will also compare our results in this
limit to earlier results for the Gaussian chain.

Eq. (\ref{cond1}) provides a general criterion for the delocalization of a
polymer with varying degree of flexibility. The next section addresses
some of the difficulties involved in treating the wormlike chain model
and presents a simplified model to treat semiflexibility. 

\section{Gaussian stiff Chains}

The wormlike chain model, though a proper description of semiflexible
polymers, suffers from its analytical intractability in a large number of
applications due to the constraint of finite extensibility, i.e., $ |{\bf
  u}(s)^{2}| = 1, \forall s \in [0,L]$. In many of the practical problems,
therefore, it is customary to relax the local constraint of unit tangent
vector with the global constraint $\left<|{\bf u}(s)^{2}|\right> = 1$ [17-24],
producing a rather simple model in which the chain Hamiltonian has a
connectivity term along with the bending energy term:
\begin{equation}
\beta {\cal H}_0 = {\eta \over 2} \int_0^L ds {\bf u}(s)^2 + {\epsilon
  \over2} \int_0^L ds  \left( \frac{\partial {\bf u}(s)} {\partial s} \right)^2
\end{equation}
The mean square end-to-end distance obtained using the above Hamiltonian is
identical with that of the wormlike chain model when $\eta = 3/2l_{p}$ and
$\epsilon = 3l_{p}/2$ [17-24]. The correlation between the unit tangent
vectors, when calculated from the above Hamiltonian, is of the form
\begin{equation}
\langle {\bf u}(s)\cdot {\bf u}(s')\rangle = \exp(-|s-s'|/l_{0}), 
\end{equation}
where $l_{0} = 2l_{p}/3$. The comparison of the above Equation with Eq. (2)
suggests that the persistence length for this approximate model for stiff
chains is smaller by a factor of $2/3$ {\cite {thiru, thomas, winkler}}. This
is not a serious concern in the context of the present problem since the
Gaussian model contains the essence of semiflexiblity, i.e., finite
extensibility, while remaining analytically tractable. Moreover, the difference between
 $l_{p}$ and $l_{0}$ becomes less important in higher dimensions \cite{thomas}. 

 
     
As before, we begin by writing down the free energy for this simplified model
using dimensional analysis or Flory-Imry-Ma arguments which includes the
Wiener stretching term:
\begin{equation}
\label{freeenergy}
\beta {\cal F} \simeq \frac{R^2}{l_{p}L} + \frac{l_pL}{R^2} - 
\left(\frac{\tilde{\Delta}}{R^3}\right)^{1/2} L.
\end{equation}
As expected, when the first term is ignored, and the free energy is minimized
with respect to $R$, one obtains $R_{\rm loc} = l_p^2/\tilde{\Delta}$.
However, the Wiener-term allows for more entropy, and it is natural to ask for
perturbation around $R_{\rm loc}$. We look for solutions of the form
\begin{equation}
\label{locsize2}
R = R_{\rm loc}(1+ \delta),
\end{equation}
where $\delta$ is a small correction. When Eq. (16) is
minimized with respect to $R$, and Eq. (\ref{locsize2}) is substituted 
in the resulting expression, $\delta$ turns out to be of the form  
\begin{equation}
\delta \propto \frac{1}{1+  (L/l_p)^2 
(\tilde{\Delta}/l_p)^4}.
\end{equation}
In the flexible limit of $l_p  < L$, Eq. (\ref{cond1})
suggests $\tilde{\Delta}/l_{p}<1$ and Eq. (\ref{locsize2}) reduces to
\begin{equation}
\label{locsize3}
R \approx R_{\rm loc} \left(1-\left(\frac{\tilde{\Delta}}{l_{p}}\right)^{4}  \right).
\end{equation}
Although these scaling arguments appear to be reasonable and in accord with
the physical intuition, the appropriate conditions and
justifications remain to be recovered from a more rigorous analysis. The next
section presents a systematic approach to this problem using Feynman
variational method in replica space. Since for delocalized phases the replica symmetry breaking is no longer essential, the idea is to see if criteria for localization and delocalization can be derived from standard replica techniques using one step replica symmetry breaking. 


\section{Variational method in replica space}

A semiflexible polymer in presence of the randomly distributed obstacles
confined in a harmonic potential is of the form
\begin{eqnarray}
\beta {\cal H} &=& \frac{\eta}{2}\int_{0}^{L} ds \left(\frac{\partial{\bf
      r}(s)}{\partial{s}}\right)^2 +  \frac{\epsilon}{2} \int_{0}^{L}
ds \left(\frac{\partial^2{\bf r}(s)}{\partial{s}^2}\right)^2\nonumber\\ 
&+& \sum_{i=1}^{N}\int_{0}^{L} ds W[{\bf r}(s)-{\bf r}_{i}] + \frac{\mu}{2}\int_{0}^{L}ds 
{\bf r}(s)^2,
\end{eqnarray}
where $\mu$ is the strength of the harmonic potential and defines the
system size; the importance of this term will be discussed
shortly.

Following the same steps as used to obtain Eq. (\ref{annealed}), the quench average over the obstacle density 
produces the following effective Hamiltonian:
\begin{eqnarray}
\label{effH}
\beta {\cal H}_{n} &=&\frac{\eta}{2} \sum_{a = 1}^{n}\int_{0}^{L} ds
\left(\frac{\partial{\bf r}_{a}(s)}{\partial{s}}\right)^2\nonumber\\
&+& \frac{\epsilon}{2} \sum_{a = 1}^{n}\int_{0}^{L} ds
\left(\frac{\partial^2{\bf r}_{a}(s)}{\partial{s}^2}\right)^2 + \frac{\mu}{2}\sum_{a = 1}^{n} \int_{0}^{L}ds {\bf r}_{a}(s)^2
\nonumber \\
&-&\frac{\Delta}{2} \sum_{a, b = 1}^{n}
\int_{0}^{L} ds \int_{0}^{L}ds^{\prime}\delta[{\bf r}_{a}(s)-{\bf
  r}_{b}(s^{\prime})].
\end{eqnarray}
It is to be noted that carrying out an average over the disorder density
produces an effective attractive interaction of strength $\Delta$
between different replicas. To estimate the average free energy we
employ the Feynman variational method in replica space \cite{Edwards,
  parisi}, which is of the form
\begin{equation}
\label{free}
n \left< F \right>  = \left< H_{n} - h_{n}\right>_{h_{n}} - \frac{1}{\beta} \int {\Pi_{a = 1}^{n}} {\cal D}[{\bf r}_{a}(s)] e^{- \beta h_n},
\end{equation}
where $h_{n}$ is the Gaussian trial Hamiltonian with variational function $\sigma(s-s^{\prime})$ characterizing the strength of the harmonic potential: 
\begin{eqnarray}
\label{trialH}
\beta h_{n} &=& \frac{\eta}{2}\sum_{a = 1}^{n}\int_{0}^{L} ds
\left(\frac{\partial{\bf r}_{a}(s)}{\partial{s}}\right)^2 \nonumber\\
&+& \frac{\epsilon}{2} \sum_{a = 1}^{n} \int_{0}^{L} ds
\left(\frac{\partial^2{\bf r}_{a}(s)}{\partial{s}^2}\right)^2 +\sum_{a = 1}^{n} \frac{\mu}{2}\int_{0}^{L}ds {\bf r}_{a}(s)^2
\nonumber \\ &-& \sum_{a, b = 1}^{n}
\int_{0}^{L}ds\int_{0}^{L} ds^{\prime} \sigma_{ab}(s-s^{\prime}){\bf
  r}_{a}(s)\cdot {\bf r}_{b}(s). 
\end{eqnarray} 
Defining the matrix of correlations by
\begin{equation}
\label{corr}
\left<{\bf r}_{a}(s)\cdot{\bf r}_{b}(s^\prime)\right>= \frac{3}{L}
\sum_{q} e^{-iq(s-s^\prime)}G_{ab}(q),
\end{equation}
and substituting into Eqs. (\ref{effH}), (\ref{free}) and
(\ref{trialH}), the free energy simplifies to   
\begin{eqnarray}
\label{free1}
\lefteqn{\frac{\beta \left< {\cal F} \right>}{L} = const + \frac{3}{2nL } \sum_{a=1}^{n} \sum_{q}
(\epsilon q^4 + \eta q^2 + \mu)G_{aa}(q)}\nonumber\\ & &  -
\frac{3}{2nL} \sum_{q}Tr \ln[{G}(q)]  \nonumber\\ & & - \frac{\Delta
  }{2n}\sum_{a,b=1}^{n} \int_{0}^{L} dx \int \frac{d{\bf
    k}}{(2\pi)^3}  e^{-\frac{{\bf k}^2}{L}
    \sum_{q} G_{aa}(q)(1- e^{-iqx})}\nonumber\\ & & - \frac{\Delta
  }{2n}\sum_{a \neq b}^{n}\int_{0}^{L} dx \int \frac{d{\bf
    k}}{(2\pi)^3}\nonumber\\
& &~~~~~~~~~~~~~\times e^{-{\frac{{\bf k}^2}{2L}\sum_{q} (G_{aa}(q) + G_{bb}(q) - 2G_{ab}
    e^{-iqx})}},
\end{eqnarray}
where delta function of Eq. (\ref{effH}) has been exponentiated and averaged
over the Gaussian random potential to produce the last two terms in
the above equation. $G_{ab}(q)$ is the propagator with respect to the trial
Hamiltonian $h_{n}$:
\begin{equation}
G_{ab}(q) = [(\epsilon q^4 + \eta q^2 +\mu)\delta_{ab} - \sigma_{ab}(q)]^{-1}. 
\end{equation}
$\sigma_{ab}(q)$, in general, is an $n \times n$ matrix, which in the case
of replica symmetric solution can be represented by a simple
variational ansatz:
\begin{equation}
\sigma_{ab}(q) = -\delta_{ab}[{\sigma}_{d}^{\prime}\delta_{q,0} + \sigma_{d}
(1-\delta_{q,0})] + \sigma (1 - \delta_{ab})\delta_{q,0},
\end{equation} 
where the diagonal variational parameters ${\sigma_{d}}^{\prime}$'s
representing the $q=0$ modes have been separated from all other modes
(represented by ${\sigma_{d}}$'s) to avoid working with infinite number of
variational parameters. The off-diagonal terms have been considered to 
depend only on the $q=0$ mode, as is conventionally done \cite{Gold1}. Translational
invariance of the variational function requires
$\sum_{b}\sigma_{ab}(q=0) = 0$, resulting in a condition
${\sigma_{d}}^{\prime}+\sigma = 0$ that leaves only two of the
variational parameters independent.

It has been argued by Cates and Ball that in the thermodynamic limit of
infinite system size the annealed and quenched average should be the same, and
an infinite Gaussian chain should always be collapsed \cite{Cates}. That this
indeed is the case has been shown by Goldshmidt,
where the $\mu \rightarrow 0$ limit corresponds to an infinite system size,
and the replica symmetric solution for a quenched random potential recovers
the results of the annealed disorder. On the contrary, when $\mu$ is taken to
be finite, the replica symmetric solution is found to be an inadequate
description of the problem, especially at high disorder strength where replica
symmetry needs to be broken \cite{Gold1}. The importance of weak replica
symmetry breaking was recognized by one of us earlier \cite{Vilgis1}, where a
special field theoretic description of polymers in strong disorder was used.
In general, for problems concerning random potentials with short range
correlations, one step replica symmetry breaking has been found to be a
sufficient description \cite{parisi}. In what follows, therefore, we employ
one step replica symmetry breaking to estimate the size of a semiflexible
polymer in a quenched random distribution of obstacles.

From Eq. (27), the mean square end-to-end distance of a polymer in terms
of the propagator $G(q)$ is given by
\begin{equation}
\left<{\bf R}^2\right> = \lim_{n\rightarrow
  0}\frac{3}{n}\sum_{a,b=1}^{n}\int \frac{dq}{2\pi} (G_{aa}(q) + G_{bb}(q) - 2G_{ab}(q)e^{-iqL}),
\end{equation}
where in the limit of large $L$ the summation over $q$ has been replaced by an
integral. Eqs. (29) and (30) when substituted into the above equation followed
by integration over $q$ produces
\begin{eqnarray} 
\label{ete1}  
\left<{\bf R}^2\right> &=& \frac{3}{(\eta + 2
  \sigma_{d}^{1/2}\epsilon^{1/2})^{1/2}~\sigma_{d}^{1/2}} 
\left(1 -\right.\nonumber\\
& & \frac{1}{\alpha_{+}^{1/2}-\alpha_{-}^{1/2}}
(\alpha_{+}^{1/2}e^{-\alpha_{-}^{1/2}\eta^{1/2}
  L/\epsilon^{1/2}}\nonumber\\
& & ~~~~~~~~~~- \left.\alpha_{-}^{1/2}e^{-\alpha_{+}^{1/2}\eta^{1/2} L/\epsilon^{1/2}})\right),
\end{eqnarray} 
where $\alpha_{\pm}= [ 1 \pm (1 - 4\epsilon\sigma_{d}/\eta^2)^{1/2}]/2$ and
 $\sigma_{d}$ is the variational parameter to be determined by
minimizing the variational free energy within one step replica
breaking scheme \cite{parisi, Gold1}. 

In replica symmetry breaking scheme, the $n\times n$
matrix $G(q)$ in the limit of $n\rightarrow 0$ is parameterized by a
function $g(q,u)$, where $u$ is a continuous variable that is defined in the interval $[0,1]$. The diagonal elements, on the other hand, are parameterized by ${\tilde{g}}(q)$ such that 
\begin{equation}
\lim_{n \rightarrow 0}\sum_{a,b=1}^{n} G_{ab}(q) = {\tilde{g}}(q) - \left<g(q)\right>,
\end{equation}
where $\left<g(q)\right> = \int_{0}^{1} du g(q,u)$. 

The idea is to compute the inverse of a hierarchical matrix $H(q) =
[(\epsilon q^4 + \eta q^2 +\mu)\delta_{ab} - \sigma_{ab}(q)]$ such
that $G(q) = H(q)^{-1}$. The matrix $H(q)$, in the limit of $n\rightarrow0$,
is similarly parameterized by the diagonal element $\tilde{h}(q)$ and
off-diagonal element $h(q,u)$ given by
\begin{equation}
h(q,u) = \epsilon q^4 + \eta q^2 + {\sigma_{d}}^{\prime}~\delta_{q,0}
+ (1 - \delta_{q,0})\sigma_{d} - \sigma(q,u),
\end{equation}
In one step replica symmetry breaking, the symmetry is broken at a point $u_{c}$, and the function is defined by
\begin{eqnarray}
\sigma(q,u) &=& \sigma_{0}~\delta_{q,0}; ~~~ u < u_{c}, \nonumber\\
          &=& \sigma_{1}~\delta_{q,0}; ~~~ u > u_{c}.
\end{eqnarray} 
Translational invariance of the variational function in the
limit of $n\rightarrow0$ for the case of one step replica symmetry
breaking requires ${\sigma_{d}}^{\prime} + u_{c}\sigma_{0} +
(1-u_{c})\sigma_{1} = 0$. The variational free energy, as given by
Eq. (\ref{free1}), in the limit of $n\rightarrow0$ is therefore of the form:
\begin{eqnarray}
\label{free2}
\lefteqn {\frac{\beta \left< {\cal F} \right>}{L} = const + \frac{3}{2} \int
\frac{dq}{2\pi} (\epsilon q^4 + \eta q^2 +
\mu){\tilde{g}}(q)}\nonumber\\& &  + \lim_{n\rightarrow0}\frac{3}{2n}
\int \frac{dq}{2\pi}Tr\ln[H(q)] \nonumber\\ & &- \frac{\Delta}{2}\int_{0}^{L} dx \int \frac{d{\bf k}}{(2\pi)^3} e^{-{\bf k}^2 \int \frac{dq}{2\pi} {{\tilde{g}}(q)(1-
  e^{-iqx})}} \nonumber \\ & & - \frac{\Delta}{2}\int_{0}^{1}
du\int_{0}^{L}dx \int \frac{d{\bf k}}{(2\pi)^3}e^{{-\bf k}^2\int \frac{dq}{2\pi} {({\tilde{g}}(q) - g(q,u) e^{-iqx})}}.\nonumber\\
\end{eqnarray}
The details of the replica symmetry breaking (RSB) method are described in
the paper by M\'{e}zard and Parisi \cite{parisi}, the appendix of which provides the explicit expressions for some of the terms introduced in the above
equation. One step replica symmetry breaking steps relevant to the
present problem are furnished in Appendix A. The free energy as given
by Eq. (A14), in the limit of large $L$ and $\mu\rightarrow0$, is dominated by
\begin{equation}
\frac{\beta \left< {\cal F} \right>}{L} =  const -
\frac{3\sigma_{d}g_{0}}{4} + \frac{3}{2 g_{0} \epsilon^{1/2}
  \sigma_{d}^{1/2}} - \tilde{\Delta}^{1/2} g_{0}^{-3/4},
\end{equation}
where $g_{0} = (\eta +
2\sigma_{d}^{1/2}\epsilon^{1/2})^{-1/2}{\sigma_{d}}^{-1/2}$ and $\tilde{\Delta} = 3\Delta|\ln\mu|/(2\pi)^{3/2}$. The minimization of the free energy with respect to $\sigma_{d}$
produces the following algebraic equation:
\begin{equation}
\label{alg}
\tilde{\Delta}^4(\eta + 2 \sigma_{d}^{1/2}\epsilon^{1/2})^7 -
\sigma_{d} = 0.
\end{equation}
That Eq. (\ref{ete1}) is a valid description in both stiff and flexible limit
can be seen as follows: In the absence of disorder (corresponding to
$\sigma_{d}\rightarrow 0$), when $L\eta^{1/2}/\epsilon^{1/2}\gg 1, ~
\left<{\bf R}^2\right> \approx 3L/\eta (\equiv 2 L l_{p}) $ (the flexible
limit); when $\eta^{1/2}L/\epsilon^{1/2} \ll~ 1, \left<{\bf R}^2\right>
\approx L^2$ (the rod limit).

\section{Results and Discussion}

\subsection{Localization-Delocalization Transition}

When $\eta = 3/2l_{p}$ and $\epsilon = 3l_{p}/2$ are substituted into Eq.
(38), it can be rewritten in the following dimensionless form:
\begin{equation}
\label{alg1}
{\Delta_{0}}^4( 1 + 3 {{\sigma}_{0}}^{1/2})^7 -
{\sigma}_{0} = 0,
\end{equation}
where ${\Delta}_{0} =\tilde{\Delta}/l_{0} $, $\sigma_{0} = \sigma_{d}{l_{0}}^3$ and $l_{0}= 2l_{p}/3$.

The range of applicability of one-step RSB solution can qualitatively be seen
from the behavior of $u_{c}$, which is defined in the interval $[0,1]$.  The
form of $u_{c}$, as given by Eq. (A11), suggests that $|\ln \mu| \leq
{L\tilde{\Delta}^{1/2}}/g_{0}^{3/4}$, where $g_{0} \equiv
l_{0}^2(1+3\sigma_{0})^{-1/2}\sigma_{0}^{-1/2}$. Since $|\ln \mu| \simeq |\ln
V|$, the above inequality implies that a chain localizes when the binding
energy is greater than the translational energy. It is to be noted that this
inequality arises as a natural condition on the validity of replica symmetry
breaking solution.  In the flexible limit of $L/l_{0} > 1$ and
$l_{0}\rightarrow 0$, Eq. (39) suggests that $\sigma_{0} \approx
(\tilde{\Delta}/l_{0})^4 $, and the above inequality yields $(l_{0}/L)^{1/2}
|\ln \mu|^{1/2} \leq {\tilde{\Delta}}/l_{0}$, the form of which is same as
suggested using simple scaling arguments after Eq. (\ref{locsize1}).

\begin{figure}
\rotatebox{270}{\includegraphics[width=8cm,height=9cm]{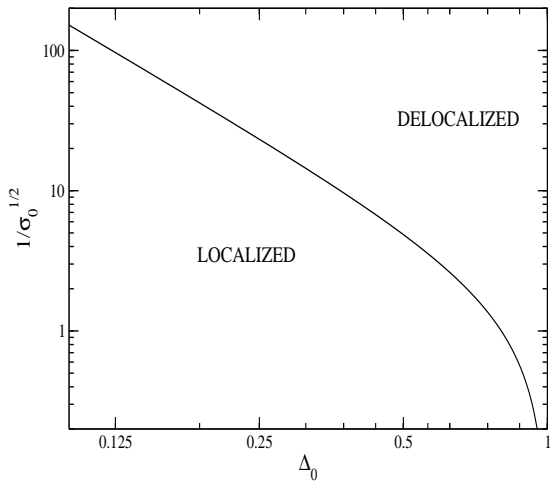}}
\caption{Numerical solution of Eq. (\ref{alg1}) representing
  localization - delocalization transition as the dimensionless variational
  parameter $\sigma_{0}$ and dimensionless disorder strength $\Delta_{0} =
  \tilde{\Delta}/l_{0}$ are varied.}
\end{figure}

Eq. (39) can be solved numerically for $\sigma_{0}$ as a function of
$\Delta_0$. The results are summarized in Fig. (1); the delocalized region
corresponds to the case where one-step RSB does not provide physically
meaningful solutions to the dimensionless variational parameter $\sigma_{0}$.
As is evident from Fig. (1), the chain is localized as long as the condition
$\tilde{\Delta}/l_{0}\leq 1$ is satisfied. The latter condition introduces an
upper cut-off on the dimensionless ratio $\tilde{\Delta}/l_{0}$; it suggests
that in the flexible limit ($0 < l_{p}/L < 1$), for a given persistence
length, there exist a critical disorder strength $\tilde{\Delta}_{c} (<
l_{0})$ above which the chain delocalizes.

The Gaussian approximation for semiflexible chains forbids us to consider
rigorously the localization of a rigid rod [17-21]. In what follows, we
restrict ourselves to the limit of intermediate stiffness $0 < l_{p}/L < 1$.
The real stiff limit of $L/l_{p} < 1$ requires a separate treatment.
\begin{figure}
\rotatebox{270}{\includegraphics[width=8cm,height=9cm]{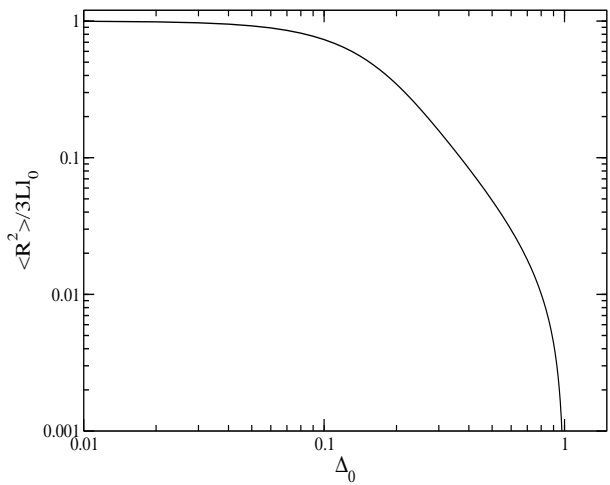}}
\caption{Log-log plot of the normalized mean square end-to-end
  distance $\left<{\bf R}^2\right>/3Ll_{p}$ vs. the dimensionless disorder
  strength $\Delta_{0} = \tilde{\Delta}/l_{0}$. The curve is the numerical
  solution of Eqs. (\ref{alg1}) and (\ref{ete2}) for the ratio
  $L/l_{0}=100.0$.}
\end{figure}

\subsection{The size of a semiflexible polymer in disorder}

In the flexible limit of $l_{0} \rightarrow 0$, Eq. (39) can be solved to
quadratic order in $\sigma_{0}$. The resulting expression is given by
\begin{equation}
\sigma_{0}^{1/2} = \frac{\Delta_{0}^4 [ 1 + (1 + 12 (1 - \Delta_{0}^4)/7\Delta_{0}^4)^{1/2}]}{2 (1 - \Delta_{0}^4)}.
\end{equation}
When $\eta = 3/2l_{p}$ and $\epsilon = 3l_{p}/2$ are substituted into Eq. (31), the resulting expression can be written in the following form:
\begin{eqnarray} 
\label{ete2}  
\left<{\bf R}^2\right> &=& \frac{3l_{0}^2}{(1 + 3 \sigma_{0}^{1/2})^{1/2}~\sigma_{0}^{1/2}} 
\left(1 -
\frac{1}{\alpha_{+}^{1/2}-\alpha_{-}^{1/2}}\right.\nonumber\\
& \times & \left. (\alpha_{+}^{1/2}e^{-2\alpha_{-}^{1/2}
  L/3l_{0}}- \alpha_{-}^{1/2}e^{-2\alpha_{+}^{1/2} L/3l_{0}})\right),
\end{eqnarray}
where $\alpha_{\pm}= [ 1 \pm (1 - 9\sigma_{0})^{1/2}]/2$. In the flexible
limit of $L/l_{0}>1$, to leading order in $l_{0}$, the above expression can be
simplified to produce the following approximate expression for the mean square
end-to-end distance:
\begin{equation}   
\left<{\bf R}^2\right> = \frac{3l_{0}^2}{\sigma_{0}^{1/2}}(1 - \exp(-\sigma_{0}^{1/2}L/l_{0})).
\end{equation}
In the limit of low disorder strength ${\Delta_{0}} \rightarrow 0$, $ R
\approx l_{0}^2/\tilde{\Delta}$, which is same as Eq. (\ref{locsize1})
obtained using scaling arguments. When the disorder strength reaches its
critical value corresponding to ${\Delta_{0}}\approx 1$, $ R \approx
l_{0}^3/\tilde{\Delta}^2$. The latter suggests that when the average
separation between the obstacles is of the order of the persistence length
($l_{\rm obs}\approx \rho_{0}^{-1/3} \approx l_{0})$, the disorder can explore
the length scale over which bending energy is stored. Since any further
increase in the disorder density amounts to high energy penalty for bending,
the polymer delocalizes above the critical density. The decrease in the size
close to $\tilde\Delta/l_{0} \simeq 1$ can therefore be attributed to the
dominance of the bending mode at this length scale. The results are summarized
quantitatively in Fig. (2), which is a numerical solution of Eqs. (\ref{alg1})
and (\ref{ete2}) and shows a sharp decrease in the size of the polymer close
to the critical disorder strength. These results are in contrast to to the
conventional studies on the localization of a Gaussian chain, which is a
fractal, and undergoes a monotonous decrease in the size with the increase in
the disorder density. In the limit of $\tilde\Delta/l_{0} < 1$, to the first
order correction in $\tilde\Delta/l_{0}$, the size is given by $R \approx
l_{p}^2(1 - \tilde\Delta^4/l_{0}^4)/\tilde{\Delta}$, which is similar to Eq.
(\ref{locsize3}). The one step replica symmetry breaking, therefore, supports 
the scaling arguments presented in Section III.
 
It is to be noted that in the above analysis we have used $l_{0}$ instead of
$l_{p}$ as the persistence length.  This is because $l_{0} = 2l_{p}/3$ appears
as the natural persistence length for Gaussian stiff chains, an issue that has
been discussed in Sec. III. In the rest of our analysis $l_{p}$ will be used
as the persistence length.

\subsection{The final state of the polymer}
\label{final}

When the polymer is able to find a suitable region among the obstacles, its
conformation must contain a rich structure. Though the results of the previous
section show a dramatic decrease in the chain size with the increase in the
disorder strength, they say nothing about several different {\it folded}
states a chain can adopt as a function of the chain rigidity $L/l_{p}$ and
dimensionless disorder strength $\tilde{\Delta}/l_{p}$. In what follows we
propose a phase diagram for several possible compact states of the localized
polymer by comparing their surface energies.

Since an average over the disorder density produces an effective attractive interaction between chain segments, a chain can find itself in several compact states depending on its stiffness. For a fully flexible chain the compact state is expected to be a globule; a
stiffer chain, on the other hand, has a tendency to wind around itself many
times to form a toroid \cite{grosberg, odijk}. In presence of obstacles, the
size of a circular toroid can be obtained by minimizing the free energy of the
form
\begin{equation}
\beta {\cal F} = \frac{l_{p} R_{s}^2}{R b^2} - \left(\frac{\tilde{\Delta}}{R^3}\right)^{1/2} L,
\end{equation}
where $b$ is the monomer size and $l_{p}$ is the length over which monomers of size $b$ are correlated. 
The first term accounts for the bending energy of a toroid
\cite{williams1}; $R_{s}$ is the radius of the torus tube, and $R$ is the
radius from the center of the hole to the center of the tube. Minimization of
the free energy with respect to $R$ yields $R \simeq l_{p}^2/\tilde{\Delta}$
and $R_{s} \simeq {\tilde{\Delta}}^{1/2}L^{1/2}b/l_{p}$, where the volume
constraint, $V = 2 \pi^2 R R_{s}^2 = L b^2$, has been used. As expected, the
minor radius $R_s$ grows at the expense of the major radius $R$ with the
decrease in the persistence length implying that for a given disorder strength
there exist a critical ratio of $L_{\rm crit}/l_{p}$ above which the chain
prefers to be in a globular state. Comparison of the surface energy penalty of
a toroid $S_{t} \simeq \gamma R R_{s}$ with respect to a sphere $S_{g} \simeq
\gamma R^2$ suggests $L_{\rm crit}/l_{p} \simeq
(l_{p}/\tilde{\Delta})^3(l_{p}/b)^2$, where $\gamma$ is the surface tension.
As expected, for high disorder strength, $\tilde{\Delta}/l_{p}$, and,
$L/l_{p}> L_{\rm crit}/l_{p}$, the chain prefers to be in a globular state. In
the opposite limit the chain adopts a toroidal state.
        
\begin{figure}
\rotatebox{270}{\includegraphics[width=8cm,height=9cm]{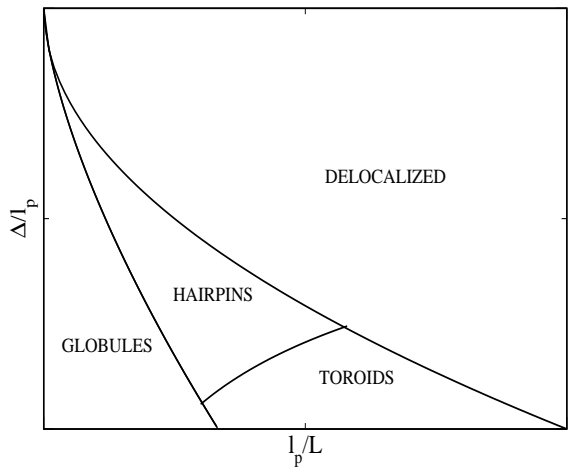}}
\caption{A schematic phase diagram based on the scaling analysis depicting
  different folded states of a localized polymer as a function of the
  dimensionless disorder strength, $\Delta/l_{p}$, and chain rigidity,
  $l_{p}/L$. The coexistence curves are the regions of comparable surface
  energies. At large $l_{p}/L$ for a given large $\Delta/l_{p}$ the chain is
 delocalized in the static sense since under these conditions (Eq. (\ref{cond1})) the mean separation between the obstacles is less than the persistence length and the chain cannot find an appropriate place in the disordered medium to sufficiently bend; it can do so only at the cost of very high bending energy.}
\end{figure}   

When the disorder strength is high, a chain can pack densely only if it
chooses a mean direction that a given distribution of obstacles imposes. In
other words, a high obstacle density favors anisotropy in structures, and it
seems unlikely that under these conditions a polymer can adopt uniformly bent
states like globules or toroids. It is physically intuitive, however, that if
the polymer is able to localize for moderate persistence lengths, it can do so
by producing many hairpins in a small available volume. The hairpins are the
abrupt reversals in the direction of the chain from the mean ordering
direction and result from the balance between the elastic penalty that favors
gradual bends, and orientational field that favors rapid reversal in the
alignment of the chain in the preferred direction. The energy costs for
forming a hairpin is given by \cite{warner, williams}
\begin{equation}
\beta U_{h} = \left( {u_{h}}{l_p} \right)^{1/2},
\end{equation}
where $u_{h}$ is the coupling constant of a (mean-field)
Maier-Saupe potential between the aligned chain segments
\cite{warner1}. Since $u_{h}$ depends on the microscpic details, a simple dimensional analysis of the above equation shows that $u_{h}$ can at most be of the order of $1/b$. The free energy for this case is given by 
\begin{equation}
\beta {\cal F} = \frac{L}{R} \left({u_{h}}{l_p} \right)^{1/2} -\left(\frac{\tilde{\Delta}}{R^3}\right)^{1/2} L ,
\end{equation}
where we have assumed that the number of hairpins $n_{h}$ must be resolved in
a self-consistent way, i.e., $n_{h} = L/R$. By minimizing the 
the free energy with respect to $R$, we find for the final state of the localized chain
\begin{equation}
R \simeq {\lambda}^2 \left(\frac{\tilde{\Delta}}{{l_p}^2}\right),
\end{equation}
where $\lambda$ represents the length of a bend given by $\lambda =
\sqrt{l_{p}/u_{h}}$, and is of the order of less than or equal to the
persistence length; $n_{\rm max} \simeq L/\lambda$ determines the maximum
number of hairpins that a chain of a fixed contour length $L$ can sustain
\cite{warner1, warner}. By comparing the surface energy cost for hairpins,
$S_{h} \simeq \gamma l_{p}R$, with that of toroids one obtains $L_{\rm
  crit}/l_{p}\simeq
(\tilde{\Delta}/l_{p})^{3}(\lambda/b)^{2}(\lambda/l_{p})^{2}$. In writing the
surface energy cost for a hairpin, we have accounted for the fact that the
average length of the chain in directions parallel to the mean ordering
direction scales as $l_{p}$. It turns out then that for high disorder strength
and $L/l_{p}> L_{\rm crit}/l_{p}$ a chain prefers to form many hairpins. In
the opposite limit toroidal state is favored. The results are summarized in
the form of a schematic phase diagram in Fig. (3). It clearly shows the
dominance of hairpins at moderate persistence length and high disorder
strength; for small and large persistence lengths the chain conformation is
dominated by uniformly bent shapes like globules and toroids respectively.

\section{Conclusions}

We interpret our results to suggest that the interplay between the two length
scales --- the persistence length of the chain and the average separation
between the obstacles --- serves to localize a semiflexible polymer in a
random distribution of obstacles. As long as the disorder density is low, the
mean separation between the obstacles is larger than the persistence length,
and the chain is localized; but as soon as the disorder explores the length
scales of the order of the persistence length the chain delocalizes since the
energy penalty for bending exceeds the energy gain in binding. On the
contrary, when the density of obstacles is kept fixed, delocalization can be
brought about by varying the temperature. While the ease of localization of a
semiflexible chain can be enhanced by increase in the temperature
(corresponding to the decrease in the persistence length), a localized
flexible chain can be delocalized by lowering the temperature. Once the
polymer is localized it adopts a highly compact folded state by forming
hairpins for moderate persistence length and high disorder strength; globules
and toroids are favored for small and large persistence lengths respectively.

Although the idea was to present a general theory for localization of
semiflexible polymers, there is one respect in which the present formalism
could be seen to have limitations, and that is the use of the Gaussian
approximation for stiff chains, which fails to adequately describe the rigid
rod limit. Nevertheless, given the range of systems that can be described by
flexible chain models, this theory holds out the possibility of addressing
several realistic problems, for instance, the immobilization of biopolymers on
surfaces with random interactions.

\acknowledgments

The authors acknowledge useful discussions with V. G. Rostiashvili. A. Dua
would like to thank R. Adhikari and S. Sukumaran for a critical reading of the
manuscript. T.A.V. thanks N. Sirkis for useful inspiration.

\appendix
\section{One-Step replica symmetry breaking}

The definitions of some of the terms introduced in
Eq. (\ref{free2}) is as follows:
\begin{eqnarray}
\lefteqn{\lim_{n\rightarrow 0}\frac{1}{n}Tr\ln[H(q)] =
\ln(\tilde{h}(q)-\left<h(q)\right>) }\nonumber\\
& &+ \frac{h(q,0)}{\tilde{h}(q)-\left<h(q)\right>}\nonumber\\ 
& & \int_{0}^{1}
\frac{du}{u^2} \ln \frac{{\tilde{h}}(q) - \left<h(q)\right> - [h](q,u)}{{\tilde{h}}(q)},\\
{\tilde{g}}(q) &=& \frac{1}{\tilde{h}(q)-\left<h(q)\right>} \left[ 1 - \frac{h(q,0)}{\tilde{h}(q)-\left<h(q)\right>}\right.\nonumber\\
& &~~~~~ - \left.\int_{0}^{1}\frac{1}{u^2}
\frac{[h](q,u)}{\tilde{h}(q)-\left<h(q)\right>-[h](q,u)}\right], \\
g(q,u) &=&
-\frac{1}{\tilde{h}(q)-\left<h(q)\right>}\left[\frac{[h](q,u)}{u(\tilde{h}(q)-\left<h(q)\right>-[h](q,u))}\right.\nonumber\\
& & ~~~~~\left. +
  \int_{0}^{\mu}\frac{d\nu}{\nu^2}\frac{[h](q,u)}{\tilde{h}(q)-\left<h(q)\right>-[h](q,u)}\right.\nonumber\\
& & ~~~~~+ \left.  \frac{h(q,0)}{\tilde{h}(q)-\left<h(q)\right>}\right]\nonumber\\
\end{eqnarray}
For the case of one step replica symmetry breaking the calculation is 
straightforward, and the explicit expressions for the above equations
are given by  
\begin{eqnarray}
{\tilde{h}}(q) - \left<h(q)\right> &=& \epsilon q^4 + \eta q^2 + \mu + \sigma_{d}(1-\delta_{q,0}),\\
{\tilde{g}}(q) &=&  \delta_{q,0}\left[ \frac{1}{\mu} +
  \frac{\sigma_{0}}{\mu^2} - \left( 1 - \frac{1}{u_{c}} \right)
  \frac{\Sigma_{1}}{\mu(\mu + \Sigma_{1})} \right]\nonumber\\
& & ~~~~~~+ \frac{(1-\delta_{q,0})}{(\epsilon q^4 + \eta q^2 +
  \sigma_{d})} \\
g(q,u<u_{c}) &=& \delta_{q,0}\frac{\sigma_{0}}{\mu^2},
\nonumber\\
 g(q,u>u_{c}) &=& \delta_{q,0}\left[ \frac{\Sigma_{1}}{u_{c} \mu(\mu +
     \Sigma_{1})} + \frac{\sigma_{0}}{\mu^2} \right], 
\end{eqnarray}
with
\begin{eqnarray}
[h](q,u) &=& 0~;~~~~~~~~~~~~~~u < u_{c},\nonumber\\
&=& -\Sigma_{1}~\delta_{q,0}~;~~~~u > u_{c},
\end{eqnarray}
where $\Sigma_{1} = u_{c}(\sigma_{1} - \sigma_{0})$. Having substituted the
above equations into Eq. (\ref{free2}) followed by integrations with respect to
the variables $q$ and ${\bf k}$, the expression for the free energy simplifies to
\begin{eqnarray}
\frac{\beta \left< {\cal F} \right>}{L} &=& const + \frac{3(\mu
  -\sigma_{d})g_{0}}{4} + \frac{3}{2g_{0}(\epsilon
  \sigma_{d})^{1/2}}\nonumber\\ 
& & + \frac{3}{2L}\left[\ln\frac{\mu}{\sigma_{d}}
  - \frac{\mu}{\sigma_{d}} + \left(
    1-\frac{1}{u_{c}}\right)\right.\nonumber\\& & ~~~~~~~\times \left.\left(\ln\frac{\mu +
      \Sigma_{1}}{\mu} - \frac{\Sigma_{1}}{\mu +
      \Sigma_{1}}\right)\right]\nonumber\\& & - \frac{\Delta L
  }{2(2\pi)^{3/2}g_{0}^{3/2}}\left( \int_{0}^{1}dx
  (g_{1}(x)^{-3/2} \right. \nonumber\\ & & ~~~~~~~ - 
 \left. g(u > u_{c})^{-3/2}) - u_{c}(g(u <
   u_{c})^{-3/2}\right.\nonumber\\& &~~~~~~~~~~ - \left.  g(u > u_{c})^{-3/2})\right),
\end{eqnarray} 
where $g_{0} = (\eta +
2\sigma_{d}^{1/2}\epsilon^{1/2})^{-1/2}{\sigma_{d}}^{-1/2}$; other
terms introduced in the above expression have the following definitions:
\begin{eqnarray}
g(u > u_{c}) &=& 1 + \frac{2}{Lg_{0}}\left( \frac{1}{\mu + \Sigma_{1}} - 
  \frac{1}{\sigma_{d}}\right),\\
g(u < u_{c}) &=& 1 + \frac{2}{Lg_{0}}\left( \frac{1}{\mu + \Sigma_{1}} - 
  \frac{1}{\sigma_{d}} \right. \nonumber\\ & &~~~~~~~~~~~~~~~ \left.+ \frac{\Sigma_{1}}{u_{c}\mu(\mu +
    \Sigma_{1})}\right),\\
g_{1}(x) &=& 1 -
\frac{1}{\alpha_{+}^{1/2}-\alpha_{-}^{1/2}}(\alpha_{+}^{1/2}e^{-\alpha_{-}^{1/2}\eta^{1/2}
  Lx/\epsilon^{1/2}}\nonumber\\
& & ~~~~~~~~~~- \alpha_{-}^{1/2}e^{-\alpha_{+}^{1/2}\eta^{1/2} Lx/\epsilon^{1/2}}),
\end{eqnarray}
where $\alpha_{\pm}= [ 1 \pm (1 - 4\epsilon\sigma_{d}/\eta^2)^{1/2}]/2$. 
The free energy can be minimized with respect to $\Sigma_{1}$ and
$u_{c}$. In the limit of large $L$ and $\mu \rightarrow
0$, the solution of the resulting equations yield
\begin{eqnarray}
\Sigma_{1} &=&   \tilde{\Delta}^{1/2} g_{0}^{-7/4}; \\
u_{c} &=&  3 g_{0}^{3/4}|\ln \mu|\tilde{\Delta}^{-1/2} L^{-1},
\end{eqnarray}
where $\tilde{\Delta} = 3\Delta|\ln\mu|/(2\pi)^{3/2}$. The substitution
of above equations into Eq. (\ref{free2}) implies the expression for the free
energy to
\begin{eqnarray}
\frac{\beta \left< {\cal F} \right>}{L} &=& 
\frac{3(\mu-\sigma_{d})g_{0}}{4} + \frac{3}{2 g_{0}
  \epsilon^{1/2}\sigma_{d}^{1/2}} - {\tilde{\Delta}^{1/2}
  g_{0}^{-3/4}}\nonumber\\ & & +
\frac{\tilde{\Delta}g_{0}^{-3/2}L}{6|\ln \mu|} \int_{0}^{1}dx
(g_{1}(x)^{-3/2} - 1)\nonumber\\ & & + \frac{\tilde{\Delta}g_{0}^{-5/2}}{2|\ln \mu|}(\tilde{\Delta}^{-1/2} g_{0}^{7/4}-\sigma_{d}^{-1}),
\end{eqnarray}
where only terms of $O(1)$ with respect to expansion in $1/L$
have been retained.

\end{document}